\documentclass{elsart}
\usepackage{ifpdf}
\usepackage{graphicx,amssymb,lineno}
\usepackage{pxfonts} 
\usepackage{multirow}
\usepackage{algorithmic}
\usepackage{algorithm}
\usepackage{colortbl}
\usepackage{hhline}

\ifpdf
\usepackage[%
  pdftitle={Instructions for use of the document class
    elsart},%
  pdfauthor={Simon Pepping},%
  pdfsubject={The preprint document class elsart},%
  pdfkeywords={instructions for use, elsart, document class},%
  pdfstartview=FitH,%
  bookmarks=true,%
  bookmarksopen=true,%
  breaklinks=true,%
  colorlinks=true,%
  linkcolor=blue,anchorcolor=blue,%
  citecolor=blue,filecolor=blue,%
  menucolor=blue,pagecolor=blue,%
  urlcolor=blue]{hyperref}
\else
\usepackage[%
  breaklinks=true,%
  colorlinks=true,%
  linkcolor=blue,anchorcolor=blue,%
  citecolor=blue,filecolor=blue,%
  menucolor=blue,pagecolor=blue,%
  urlcolor=blue]{hyperref}
\fi

\makeatletter
\def\elsartstyle{%
    \def\normalsize{\@setfontsize\normalsize\@xiipt{14.5}}
    \def\small{\@setfontsize\small\@xipt{13.6}}
    \let\footnotesize=\small
    \def\large{\@setfontsize\large\@xivpt{18}}
    \def\Large{\@setfontsize\Large\@xviipt{22}}
    \skip\@mpfootins = 18\p@ \@plus 2\p@
    \normalsize
}
\@ifundefined{square}{}{\let\Box\square}
\makeatother

\begin{document}

\begin{frontmatter}
\title{Discovering More Accurate Frequent Web Usage Patterns}

\author[ub]{Murat Ali Bayir},\texttt{ }\ead{mbayir@cse.buffalo.edu}\author[metu]{Ismail Hakki Toroslu},\texttt{ }\ead{toroslu@ceng.metu.edu.tr}\author[metu]{Ahmet Cosar},\texttt{ }\ead{cosar@ceng.metu.edu.tr}
\author[agm]{Guven Fidan} \ead{guven.fidan@agmlab.com}

\address[ub]{Department of Computer Science and Engineering, 
University at Buffalo, SUNY, 14260, Buffalo, NY, USA}
\address[metu]{Department of Computer Engineering, Middle
East Technical University, 06531, Ankara, Turkey}
\address[agm]{AGMLAB Information Technologies, CyberPark 
Cyberplaza, Bilkent, 06800, Ankara, Turkey}

\begin{abstract}
Web usage mining is a type of web mining, which exploits data mining techniques to discover valuable information from navigation behavior of World Wide Web users. The first phase of web usage mining is the data processing phase, which includes the session reconstruction operation from server logs. Session reconstruction success directly affects the quality of the frequent patterns discovered in the next phase. In reactive web usage mining techniques, the source data is web server logs and the topology of the web pages served by the web server domain. Other kinds of information collected during the interactive browsing of web site by user, such as cookies or web logs containing similar information, are not used. The next phase of web usage mining is discovering frequent user navigation patterns. In this phase, pattern discovery methods are applied on the reconstructed sessions obtained in the first phase in order to discover frequent user patterns. In this paper, we propose a frequent web usage pattern discovery method that can be applied after session reconstruction phase. In order to compare accuracy performance of session reconstruction phase and pattern discovery phase, we have used an agent simulator, which models behavior of web users and generates web user navigation as well as the log data kept by the web server.
\end{abstract}

\begin{keyword}
Web usage mining, session reconstruction, apriori technique, agent simulator and web topology
\end{keyword}
\end{frontmatter}

\section{Introduction}

The goal in web mining \cite{CooleyMS97} is to discover and retrieve useful and interesting patterns from a large dataset. The source data for web mining contains various information sources in different formats. Web usage mining (WUM) \cite{SrivastavaCDT00} is a new research area which can be defined as a process of applying data mining techniques to discover interesting patterns from web usage data. Web usage mining provides information for better understanding of server needs and web domain design requirements of web-based applications. Web usage data contains information about the identity or origin of web users with their browsing behaviors in a web domain. Web pre-fetching \cite{Pitkow99,SchechterKS98}, link prediction \cite{Gunduz03,Frias-MartinezK02,Nanopoulous01}, site reorganization \cite{Spiliopoulou00,SrikantY01} and web personalization \cite{MobasherCS00,MobasherDLN02,NasraouiK02,PierrakosPPS03} are common applications of WUM. 

WUM data contains users' navigation behaviors on the web. Navigation among web pages by using hyperlinks is the most common action of the web user. Two web pages can be accepted as related to each other if both of them are accessed in the same user session such that the first page accessed is connected to the second one with a hyperlink. In order to support the claim about two pages being related, such accesses must occur several times. Therefore, in WUM, first user navigation sessions must be reconstructed from server access logs, and then, frequent patterns in these sessions must be searched.

Reconstruction of accurate user sessions from server access logs is a challenging task since access log protocol is stateless and connectionless. For reactive strategies, all users behind a proxy server will have the same IP number also. Moreover, caching performed by the clients' browsers and proxy servers will affect the web log data. These problems can be handled by proactive strategies by using cookies and/or java applets. However, these solutions could have been disabled by some clients for security/privacy concerns. In such cases proactive strategies become unusable.
Reactive session reconstruction and proactive session reconstruction strategies use different data sources. Proactive strategies \cite{FuS02,ShahabiK03} uses raw data collected during run-time which is usually supported by dynamic server pages. Whereas in reactive strategies \cite{CooleyMS99,CooleyTS99,SpiliopoulouF98}, server logs are main data source. Reactive strategies are mostly applied on static web pages. Because the content of dynamic web pages changes in time, it is difficult to predict the relationship between web pages and obtain meaningful navigation path patterns. Therefore we restrict our work to static web pages. As it is stated above, server logs are the main data source of reactive strategies. The information required to obtain session information are user's IP address, access date and time, and the URL of the page accessed. These three attributes are included in common log format \footnote{http://www.w3.org/Daemon/User/Config/Logging.html\#common-logfile-format}. 

There are several previous works related to mining web access patterns \cite{CooleyTS99,Gaul00,PeiHMZ00,SrivastavaCDT00}. We use modified apriori technique adapted for sequence discovery for discovering frequent access paths. This idea is not new \cite{AgrawalS95,Gaul00,PeiHMZ00}, however, to the best of our knowledge, the use of web topology for extending the large itemsets through iterations of the apriori technique is novel. In this paper, not only we show that the discovery of frequent maximal navigation patterns from already reconstructed patterns utilizing the web topology can be done very easily, but we also show that the accuracy of the discovered frequent patterns is much higher than the accuracy of the reconstructed sessions. Therefore, it is worthwhile to make extra effort to increase the accuracy of the reconstructed sessions.

The main aim of our work is to discover frequent user session patterns. The results of this work can be used in applications such as web pre-fetching. The problem of which page will be requested from the current page can be solved by applying some statistical methods to frequent pattern set generated by our method. In addition to web perfecting, link topology can be modified by examining frequent patterns. Reaching popular pages in frequent patterns can be made easier by changing link topology. Length of the most frequent navigation paths can be decreased by analyzing frequent patterns discovered by our method. By changing the link topology, web users' searches for target pages becomes easier.  

This paper is organized as follows. The next section is dedicated to session reconstruction operation. It first summarizes previously used reactive heuristics, and a recently proposed heuristics. After that, it introduces the agent simulator that was used to evaluate different session reconstruction heuristics, and finally it experimentally evaluates the accuracy of the first phase. Section 3 discusses pattern discovery from the reconstructed sessions, firstly by introducing a modified apriori technique used for pattern discovery, and then it analyzes the performance of pattern discovery phase. Finally, we give our conclusions.

\section{Session Reconstruction}
\subsection{Previous Heuristics}

Previous Reactive session reconstruction heuristics \cite{SpiliopoulouMBN03} use page access timestamps and navigation information of the users. Time oriented heuristics \cite{CooleyMS99,SpiliopoulouF98} are based on time limitations on total session time or page-stay time. In the first type, total time of the session can not be greater than predefined threshold. In the second type, predefined threshold is used for checking page-stay time. Time oriented heuristics lack path information since they do not consider page connectivity. 

Navigation-oriented approach \cite{CooleyMS99,CooleyTS99} takes web topology in graph format. It considers webpage connectivity, however, it is not necessary to have hyperlink between two consecutive pages. In case of any missing link, backward browser movements are inserted if one of the previously accessed pages refers to new page. In navigation-oriented heuristics artificially inserted links with backward browser movements is a major problem, since although the rest of the session always corresponds to forward movements in web topology graph. It is difficult to interpret these patterns. Sequential pages accessed from server side can not be extracted. In addition, extra backward movements makes sessions longer. Also there is no time limitation, for a client which has access set in very different time. The length of the session becomes very long.

\subsection{Smart-SRA}

Smart-SRA \cite{Bayir06-1,Bayir06-2} is new method proposed by us for solving deficiencies of time and navigation oriented heuristics. Smart-SRA produces sessions containing sequential pages accessed from server-side satisfying following rules:

\textit{Timestamp Ordering Rule: }

\begin{itemize}
		\item $\forall$ $i:$ 1$\le$i$<$n, Timestamp($\texttt{P}_{\texttt{i}}$) $<$ Timestamp($\texttt{P}_{\texttt{i+1}}$) 
		\item $\forall$ $i:$ 1$\le$i$<$n, Timestamp($\texttt{P}_{\texttt{i+1}}$) $-$ Timestamp($\texttt{P}_{\texttt{i}}$) $\le$ \texttt{ } $\rho$ (page stay time) 
	\item	Timestamp($\texttt{P}_{\texttt{n}}$) - Timestamp($\texttt{P}_{\texttt{1}}$) $<$ $\delta$ (session duration time).
\end{itemize}

\textit{Topology Rule:}  

\begin{itemize}
	\item $\forall$ $i:$ 1$\le$i$<$n, there is a hyperlink from $\texttt{P}_{\texttt{i}}$ to $\texttt{P}_{\texttt{i+1}}$
\end{itemize}

Smart-SRA uses page-stay and session duration rules of time-oriented heuristics. It uses topology rule as in navigation-oriented heuristics. It can be accepted as improved version of combined time and navigation oriented heuristics since it performs path completion and separation more intelligently. Smart-SRA composed of two phases. 
In the first phase of Smart-SRA, time criteria (page-stay and session duration) are applied for generating shorter sequences from raw input. In the second phase, maximal sub-sessions are generated from sequences generated in the first phase in a way that each consecutive page satisfies topology rule. Session duration time is also guaranteed by the first phase. However, page stay time should be controlled since consecutive web page pair generated in the first phase can be changed in second phase from the set of pages satisfying session duration time.  

In the first phase of Smart-SRA, time criteria (page-stay and session duration) are applied for generating shorter sequences from raw input. In the second phase, maximal sub-sessions are generated from sequences generated in the first phase in a way that each consecutive page satisfies topology rule. Session duration time is also guaranteed by the first phase. However, page stay time should be controlled since consecutive web page pair generated in the first phase can be changed in second phase from the set of pages satisfying session duration time.  

\begin{figure}
	\centering
		\includegraphics{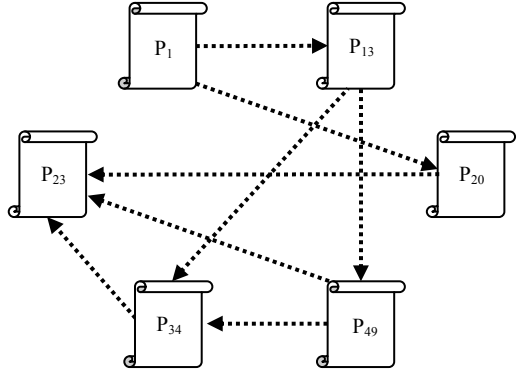}
	\caption{\small An example web site topology graph \label{fig:deneme}}
\end{figure}

The second phase adds referrer constraints of the topology rule by eliminating the need for inserting backward browser moves. This is achieved by repeating the following steps until all pages in a candidate session obtained after the first phase have been processed:

\begin{enumerate}
	\item The web pages without any referrers are determined in the candidate session from the web topology.
\item These pages are removed from the candidate session.
\item They are appended to the previously constructed sessions, if there is a hyperlink from the last page of a session to new web pages.
\end{enumerate}

Considering the web topology given in Figure 1, for the candidate session [$\texttt{P}_{\texttt{1}}$, $\texttt{P}_{\texttt{20}}$, $\texttt{P}_{\texttt{23}}$, $\texttt{P}_{\texttt{13}}$, $\texttt{P}_{\texttt{34}}$] obtained after the first phase, Smart-SRA discovers the sessions [$\texttt{P}_{\texttt{1}}$, $\texttt{P}_{\texttt{20}}$, $\texttt{P}_{\texttt{23}}$] and [$\texttt{P}_{\texttt{1}}$, $\texttt{P}_{\texttt{13}}$, $\texttt{P}_{\texttt{34}}$]. 

\subsection{Agent Simulator}

It is not possible to use web server supplied real user navigation data for evaluating and comparing different web user session reconstruction heuristics since all of the actual user requests cannot be captured by processing server side access logs. Especially the sessions containing access requests served from a client's and/or proxy's local cache cannot be known or predicted by a web server. Therefore, we have developed an agent simulator that generates web agent requests by simulating an actual web user \cite{Bayir06-1, Bayir06-2}. 

Our agent simulator first randomly generates a typical web site topology and then simulates a user agent that accesses this domain from its client site and navigates (randomly) in this domain like a real user.  In this way, we will have full knowledge about the sessions beforehand, and later when we use a heuristic to process user access log data to discover the sessions, we can evaluate how successful that heuristic was in reconstructing the known sessions. While generating a session, our agent simulator eliminates web user navigations provided via a client's local cache. Since the simulator knows the full navigation history at the client side, it can determine navigation requests that are served by the web server, and those are served from the client/proxy cache. Also, our agent simulator knows which page is the actual referrer (a page with a hyperlink to the accessed page, and this new page is accessed by following this link) of any page requested from server. 

Agent simulator produces an access log file at server side containing page requests whose pages are provided by the web server. The sessions discovered by a heuristics are compared with the original complete session file. For example, consider an agent with complete page sequences of [$\texttt{P}_{\texttt{1}}$, $\texttt{P}_{\texttt{20}}$, $\texttt{P}_{\texttt{23}}$] and [$\texttt{P}_{\texttt{1}}$, $\texttt{P}_{\texttt{13}}$, $\texttt{P}_{\texttt{34}}$] generated by the agent simulator, which correspond to the real sessions. However, in the web server log, this sequence may appear as [$\texttt{P}_{\texttt{1}}$, $\texttt{P}_{\texttt{20}}$, $\texttt{P}_{\texttt{23}}$, $\texttt{P}_{\texttt{13}}$, $\texttt{P}_{\texttt{34}}$], since the browser of the client can provide the movement from $\texttt{P}_{\texttt{23}}$ to $\texttt{P}_{\texttt{13}}$ through $\texttt{P}_{\texttt{1}}$ using its local cache, which means the second request for page $\texttt{P}_{\texttt{1}}$ will not be sent to the web server. In this example, our agent simulator generates an agent acting as a web user who requests pages $\texttt{P}_{\texttt{1}}$, $\texttt{P}_{\texttt{20}}$, $\texttt{P}_{\texttt{23}}$ consecutively, and then, returns backward to $\texttt{P}_{\texttt{1}}$, and requests page $\texttt{P}_{\texttt{13}}$. Therefore, our agent simulator knows that the actual referrer of $\texttt{P}_{\texttt{13}}$ is $\texttt{P}_{\texttt{1}}$. Finally, user agent requests page $\texttt{P}_{\texttt{34}}$ from page $\texttt{P}_{\texttt{13}}$, and thus, the agent simulator generates a session [$\texttt{P}_{\texttt{1}}$, $\texttt{P}_{\texttt{13}}$, $\texttt{P}_{\texttt{34}}$]. Heuristics used to reconstruct user sessions are run on the server side log data, and they construct candidate session sequences. These candidate sequences are compared to the real session sequences in order to determine the accuracy of the heuristics.  

An important feature of our agent simulator is its ability to model dynamic behaviors of a web agent. It simulates four basic behaviors of a web user. These behaviors can be used to construct more complex navigation behaviors in a single session. These four basic behaviors constructing complex navigations are given below:

\begin{enumerate}
	\item A Web user can start session with any one of the possible entry pages of a web site. This behavior includes new page which is not requested by any other previous page accessed from the same domain in near-time
\item A Web user can select the next page having a link from the most recently accessed page. 
\item A Web user can press the back button one more time and thus selects as the next page a page having a link from any one of the previously browsed pages (i.e., pages accessed before the most recently accessed one).
\item A Web user can terminate his/her session.
\end{enumerate}

Agent simulator also uses time considerations while simulating the behaviors described above. In the second and the third behaviors, the time difference between two consecutive page requests is smaller than 10 minutes. Also, in these behaviors, time differences of access time of the next page and the current page will have a normal distribution.  In addition, the median value for a page stay time is taken as 2.12 minutes (from \cite{SpiliopoulouMBN03}), and the standard deviation is taken as 0.5 minutes. The generated time differences set for each type of these behaviors constitute a normal distribution.

Four primitive basic behaviors given above are implemented in our agent simulator. Also, the following parameters are used for simulating navigation behavior of a web user. 

\textbf{Session Termination Probability (STP):} STP is increased as the length of a user session increases. The probability of terminating a session at the $n^{th}$ request is defined as $(1 - (1-STP)^{n})$. 

\textbf{Link from Previous pages Probability (LPP):} LPP is the probability of referring next page from one of the previously accessed pages except the most recently accessed one. This parameter is used to allow the generation of backward movements from browser. 

\textbf{New Initial page Probability (NIP):} NIP represents the probability of selecting one of the starting pages of a web site during the navigation, thus starting a new session.

\subsection{Performance Evaluation of Session Reconstruction Phase}

The most important performance criterion of the session reconstruction heuristics is the accuracy of the constructed sessions. Agent simulator can be used to measure the accuracy of the session reconstruction phase. We can simply define the accuracy of a heuristic is as the ratio of correctly reconstructed sessions over the number of real sessions generated by the agent simulator. 

A reconstructed session is correct if it captures a real session. We assume that a session H, reconstructed by a heuristic, captures a real session R, if R occurs as a subsequence of H. A session P with length n is a sub-session of a session S with length m (denoted as $P$ $\sqsubset$ $S$) if there is an index k of S, such that, 1=k=m and k+n-1=m, that satisfies the following:

            $\texttt{S}_{\texttt{k}}$ = $\texttt{P}_{\texttt{1}}$, $\texttt{S}_{\texttt{k+1}}$ = $\texttt{P}_{\texttt{2}}$, $\texttt{S}_{\texttt{k+2}}$ = $\texttt{P}_{\texttt{3}}$ \ldots    $\texttt{S}_{\texttt{k+n-1}}$ = $\texttt{P}_{\texttt{n}}$

For example, if R = [$\texttt{P}_{\texttt{1}}$, $\texttt{P}_{\texttt{3}}$, $\texttt{P}_{\texttt{5}}$] and H = [$\texttt{P}_{\texttt{9}}$, $\texttt{P}_{\texttt{1}}$, $\texttt{P}_{\texttt{3}}$, $\texttt{P}_{\texttt{5}}$, $\texttt{P}_{\texttt{8}}$], then, $R$ $\sqsubset$ $H$ since $\texttt{P}_{\texttt{1}}$, $\texttt{P}_{\texttt{3}}$ and $\texttt{P}_{\texttt{5}}$ are elements of H and they are all in the same order. On the other hand, if H = [$\texttt{P}_{\texttt{1}}$, $\texttt{P}_{\texttt{9}}$, $\texttt{P}_{\texttt{3}}$, $\texttt{P}_{\texttt{5}}$, $\texttt{P}_{\texttt{8}}$], then, $R$ $\nsqsubset$ $H$, because $\texttt{P}_{\texttt{9}}$ interrupts R in H. Searching real sessions in candidate sessions produced by heuristics can be done by using a simple algorithm adopted from an ordinary string searching algorithm.

Our agent simulator first generates a web domain, and then it produces simulated sessions and a corresponding web log file containing client requests for web pages. Then, a reconstruction heuristic processes this log file and generates candidate sessions. After that, the accuracy of the heuristics can be determined by using the reconstructed sessions and original simulated sessions.

As mentioned above, the accuracy of session reconstruction heuristics can be calculated with respect to 3 parameters, namely STP, LPP, and NIP. For evaluating the accuracy performance of different heuristics, random web sites and web agent navigations are generated by using the parameters given in Table 1. The number of web pages in a web site and the average number of out degrees of the pages (number of links from one page to other pages in the same site) are taken from \footnote{http://www.sims.berkeley.edu/research/projects/how-much-info/internet/rawdata.html}. Varying values of the three parameters defined in the previous section, namely STP, LPP, and NIP, are used for comparing the performances of the heuristics. 

In our experiments, we have fixed two of these parameters and obtained performance results for the third parameter. Therefore, three sets of experiments are performed. In the first experiment, LPP and NIP are fixed as 30\%, and STP is varying from 1\% to 20\%. In the second experiment, LPP is varying from 1\% to 90\% and STP is fixed as 5\% and NIP is fixed as 30\%. Similarly, in the third experiment, NIP is varying and STP and LPP are fixed as 5\% and 30\% respectively.

In the first experiment, increase in STP leads to sessions with fewer pages. The accuracy is higher for shorter sessions. If the navigation is affected by LPP and NIP, then, the session becomes more complex. If there is no return back to an already visited page and there is no new initial page, then, the session becomes simple and it can easily be captured. So, increasing NIP and LPP decreases the accuracy performance in contrast to STP. The accuracies of 4 heuristics (limited total session time: TO1, limited page stay time: TO2, navigation oriented: NO and Smart-SRA: SSRA) for various parameters are given in Figures 2, 3 and 4. As it can be seen from these figures Smart-SRA outperforms other previous heuristics (see \cite{Bayir06-1,Bayir06-2} for more details).

\begin{table}
\caption{Agent Simulators parameters}
\begin{tabular}{|c|c|} \hline
{\textbf{Parameter}} & {\textbf{Range}} \\ \hline
{Average Number of web pages (nodes) in topology} & {300} \\ \hline
{Average number of outdegree} & {15} \\ \hline
{Average number of page stay time} & {$2.2$min} \\ \hline

{Deviation for page stay time} & {$0.5$min} \\ \hline
{Number of agents} & {10000} \\ \hline
{Session Termination Probability (STP)} & {Fixed: 5\%, Varying:[1\%,20\%]} \\ \hline
{Link From Previous Page  probability (LPP)} & {Fixed: 30\%, Varying:[0\%,90\%]} \\ \hline
{New Initial Page probability  (NIP)} & {Fixed: 30\%, Varying:[0\%,90\%]} \\ \hline
\end{tabular}
\end{table}

\begin{figure}
	\centering
		\includegraphics{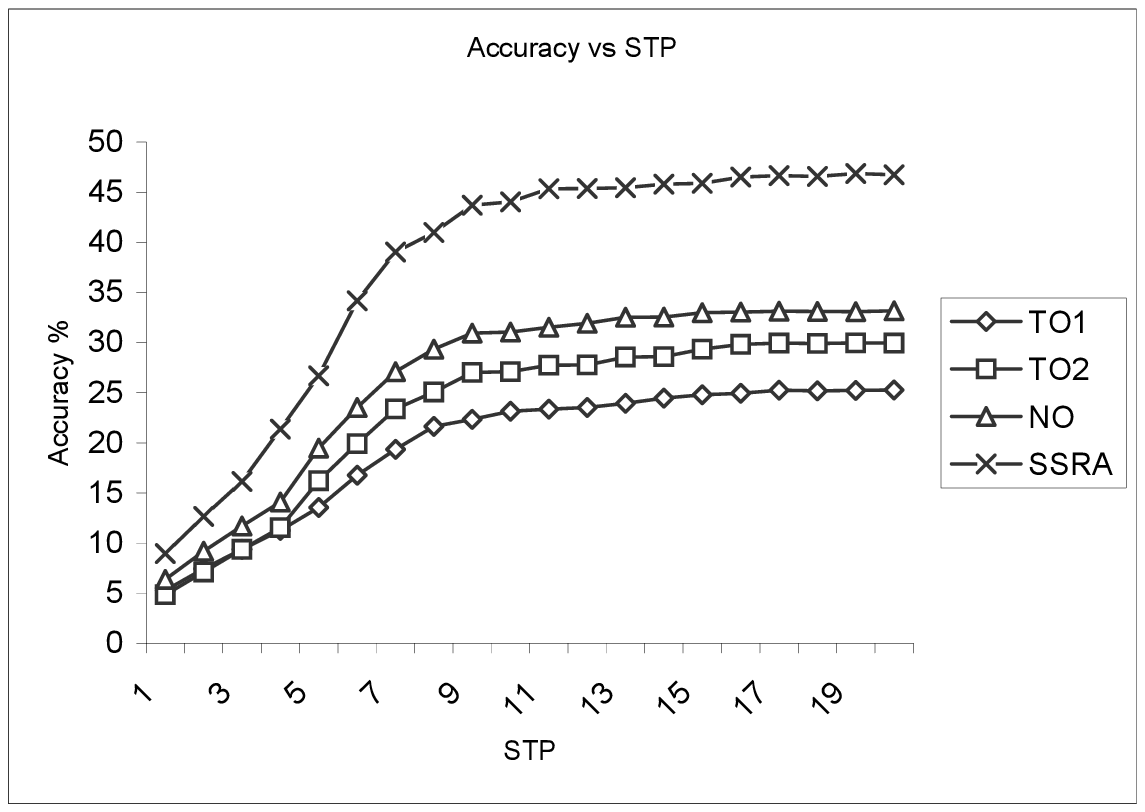}
	\caption{Reconstructed session accuracy for varying STP}
\end{figure}

\begin{figure}
	\centering
		\includegraphics{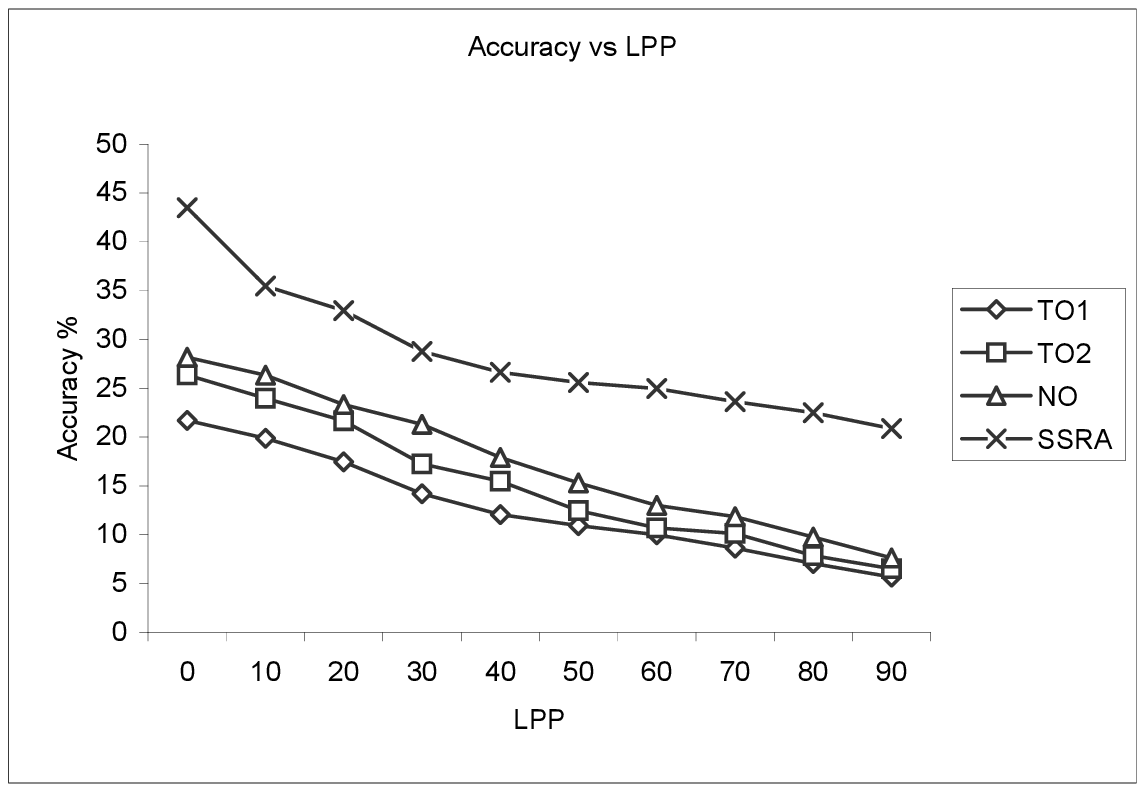}
	\caption{Reconstructed session accuracy for varying LPP}
\end{figure}

\begin{figure}
	\centering
		\includegraphics{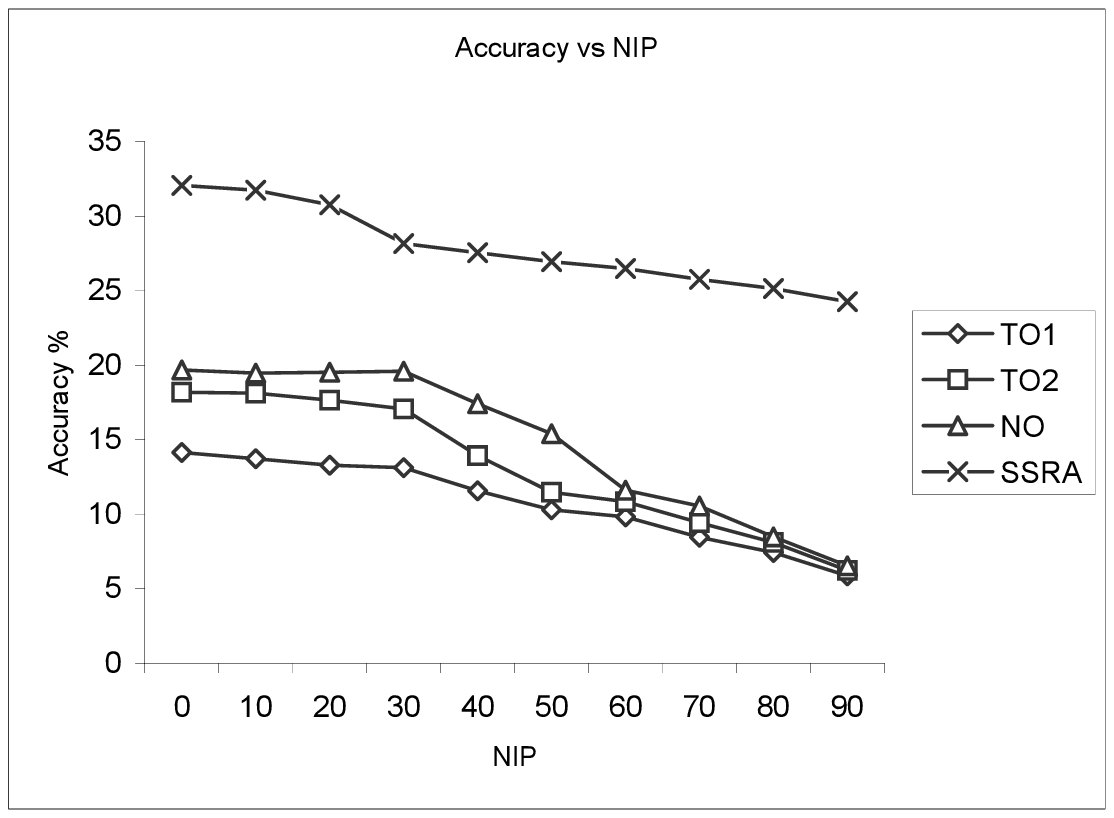}
	\caption{Reconstructed session accuracy for varying NIP}
\end{figure}

\section{Discovering Patterns}
\subsection{Sequential Apriori Technique}

A modified version of the classical apriori \cite{AgrawalS94} technique was used for discovering the frequent user access patterns from the reconstructed maximal sessions.  Unlike the ordinary large itemset discovery problem, in the user web access pattern discovery problem, consecutive pages in the discovered pattern should also appear in consecutive positions in the reconstructed sessions supporting the pattern. Therefore, frequent web access patterns can be obtained from reconstructed sessions by using a more efficient and simplified version of apriori technique.
A session S supports a pattern P if and only if P is a subsequence of S ($P$ $\sqsubset$ $S$). We call all the sessions supporting a pattern as its support set. That is, a reconstructed session S $\in$ SupportSet(P)  if  $P$ $\sqsubset$ $S$. 

\textbf{Sequential AprioriAll Algorithm (Algorithm 3):} In the beginning, each page with sufficient support forms a length-1 supported pattern. Then, in the main step, for each k value greater than 1 and up to the maximum reconstructed session length, supported patterns (patterns satisfying the support condition) with length k+1 are constructed by using the supported patterns with length k and length 1 as follows:

\begin{itemize}
	\item If the last page of the length-k pattern has a link to the page of the length-1 pattern, then by appending that page length-k+1 candidate pattern is generated.
\item If the support of the length-k+1 pattern is greater than the required support, it becomes a supported pattern. In addition, the new length-k+1 pattern becomes maximal, and the extended length-k pattern and the appended length-1 pattern become non-maximal.
\item If the length-k pattern obtained from the new length-(k+1) pattern by dropping its first element was marked as maximal in the previous iteration, it also becomes non-maximal.
\item At some k value, if no new supported pattern is constructed, the iteration halts.
\end{itemize}

Notice that in the sequential apriori algorithm, the patterns with length-k are joined with the patterns with length-1 by considering the topology rule. This step significantly eliminates many unnecessary candidate patterns before even calculating their supports, and thus increases the performance drastically. In addition, since the definition of the support automatically controls the timestamp ordering rule with the sub-session check, all discovered patterns will satisfy both the topology and the timestamp rules, which are very important in web usage mining. 

\begin{algorithm}
\caption{Sequential Apriori}
\begin{algorithmic}[1]

	\STATE \textbf{input:} Minimum support frequency: $\delta$
	\STATE Reconstructed sessions: S
	\STATE Topology information as matrix: Link
	\STATE The Set of Web Pages: P
	\STATE \textbf{output:} Set of maximal frequent patterns: Max
	
	\STATE \textbf{procedure} \underline{sequentialApriori} ($\delta$, S, Link, P)
	\STATE \texttt{ } $\texttt{L}_{\texttt{1}}$ := \{\} \COMMENT{Set of frequent length-1 patterns}
  \STATE \texttt{ } \textbf{for } i:=1 \textbf{to} $|$P$|$ \textbf{do}
  \STATE \hspace{6 mm}  $\texttt{L}_{\texttt{1}}$ := $\texttt{L}_{\texttt{1}}$ U \{[$\texttt{P}_{\texttt{i}}$] $|$ if Support([$\texttt{P}_{\texttt{i}}$],S) $>$ $\delta$ \}
	\STATE \texttt{ }   \textbf{for} k=1 \textbf{to} N-1 \textbf{do}
  \STATE \hspace{6 mm}    \textbf{if} $\texttt{L}_{\texttt{k}}$ = {} \textbf{then}
	\STATE \hspace{8 mm}    \textbf{Halt}
	\STATE \hspace{6 mm}    \textbf{else}
	\STATE \hspace{8 mm}    $\texttt{L}_{\texttt{k+1}}$ := \{\}
	\STATE \hspace{8 mm}\textbf{for each} $\texttt{I}_{\texttt{i}}$ $\in$  $\texttt{L}_{\texttt{k}}$ 
	\STATE \hspace{10 mm}\textbf{for each} $\texttt{P}_{\texttt{j}}$ $\in$ P
	\STATE \hspace{12 mm}\textbf{if} Link[LastPage($\texttt{I}_{\texttt{i}}$), $\texttt{P}_{\texttt{j}}$] = true \textbf{then}  
	\STATE \hspace{14 mm}T := $\texttt{I}_{\texttt{i}}$ $\bullet$ $\texttt{P}_{\texttt{j}}$  //Append $\texttt{P}_{\texttt{j}}$ to $\texttt{I}_{\texttt{i}}$
	\STATE \hspace{14 mm}\textbf{if} Support(T,S) $>$    \textbf{then}
	\STATE \hspace{18 mm}\texttt{T}.maximal := TRUE  
	\STATE \hspace{18 mm}$\texttt{I}_{\texttt{i}}$.maximal := FALSE //since extended
	\STATE \hspace{18 mm}V:=[$\texttt{T}_{\texttt{2}}$, $\texttt{T}_{\texttt{3}}$,\ldots, $\texttt{T}_{\texttt{$|$T$|$}}$]  		
	\COMMENT{drop first element}
	\STATE \hspace{18 mm}\textbf{if} V $\in$  $\texttt{L}_{\texttt{k}}$ \textbf{then}  
	\STATE \hspace{22 mm}V.maximal := FALSE            
	\STATE \hspace{22 mm}$\texttt{L}_{\texttt{k+1}}$ := $\texttt{L}_{\texttt{k+1}}$ U \{T\}
	\STATE \hspace{18 mm}\textbf{end if}
	\STATE \hspace{14 mm}\textbf{end if}
	\STATE \hspace{12 mm} \textbf{end if}
	\STATE \hspace{10 mm} \textbf{end for each}
	\STATE \hspace{8 mm} \textbf{end for each}
	\STATE \hspace{6 mm} \textbf{end if}
	\STATE \texttt{ }\textbf{ end for}
	\STATE \texttt{ }Max := \{\}
	\STATE \texttt{ }\textbf{for} k :=1 \textbf{to} N-1 \textbf{do}
	\STATE \hspace{6 mm}Max := Max U \{S $|$ S $\in$ $\texttt{L}_{\texttt{k}}$ and S.maximal = true\}
	\STATE \texttt{ }\textbf{end for}
	\STATE \textbf{end procedure}

\end{algorithmic}
\end{algorithm}

An auxiliary function Support (I:Pattern,S) determines whether a given pattern has sufficient support from the given set of reconstructed user sessions. Support of a pattern I is defined as a ratio between the numbers of reconstructed sessions supporting the pattern I, the number of all sessions.

\begin{eqnarray}
Support(I,S) = \frac{{|\{ S_i |\forall i\,and\,I\,is\,substring\,of\,S_i \} |}}{{|S|}}
\end{eqnarray}

Let the list of sessions in Table 2 be generated by some session reconstruction heuristic from the server logs. Let  $\delta=0.40$ be taken as minimum support for the Sequential Apriori algorithm. Then, the execution of the sequential apriori technique will generate patterns with their frequencies in three iterations as it is shown in Table 3. In this table, the patterns shown in gray areas are eliminated due to their insufficient support. Since at iteration 4, there are no remaining frequent patterns, the algorithm stops. The maximal frequent patterns are shown in bold in Table 3. The only maximal pattern is $[P_{1}, P_{13}, P_{49}]$ with support 0.40.

\begin{table}
\centering
\caption{Reconstructed Sessions Database}
\begin{tabular}{|c|c|} \hline
\textbf{Session Id} & {\textbf{Session}} \\ \hline
{1}& {[$P_{1}, P_{13}, P_{49}, P_{23}$]} \\ \hline
{2}& {[$P_{1}, P_{13}, P_{34}, P_{23}$]} \\ \hline
{3}& {[$P_{1}, P_{13}, P_{49}$]} \\ \hline
{4}& {[$P_{1}, P_{20}, P_{23}$]} \\ \hline
{5}& {[$P_{13},P_{49}$]} \\ \hline
\end{tabular}
\end{table}

\begin{table}
\centering
\caption{Patterns Generated at each Iteration}
\begin{tabular}{|c|c|c|} \hline
\textbf{\textbf{Step}} & {\textbf{Patterns}} & {\textbf{Frequencies}} \\ \hline
\multirow{3}{*}{\small 1} & {\small \{$[P_{1}], [P_{13}], [P_{23}],$} & {\small \{0.80, 0.80, 0.60, 0.60\}}  \\ 
									 & {\small $[P_{49}]$\}} & {\small $\geq 0.40$} \\ \hhline{|~|-|-|}
									 & {\cellcolor[gray]{.9} \small \{$[P_{20}],[P_{34}]$\}} & {\cellcolor[gray]{.9} \small \{0.20, 0.20\} $<$ 0.40}  \\ \hline

\multirow{2}{*}{\small 2} & {\small \{$[P_{1}, P_{13}], [P_{13}, P_{49}]$\}} & {\small \{0.60, 0.60\} $\geq 0.40$} \\  \hhline{|~|-|-|}
					&{\cellcolor[gray]{.9} \small \{$[P_{49}, P_{23}]$\}} & {\cellcolor[gray]{.9} \small \{0.20\} $<$ 0.40}  \\  \hline

\multirow{2}{*}{\small 3} & {\small \boldmath \{$[P_{1}, P_{13}, P_{49}]$\}} & {\small \{0.40\} $\geq 0.40$} \\ \hhline{|~|-|-|}
									 & {\cellcolor[gray]{.9} \small  \{$[P_{13}, P_{49}, P_{23}]$\}} & {\cellcolor[gray]{.9} \small \{0.20\} $<$ 0.40}  \\ \hline
							
\end{tabular}
\end{table}

\subsection{Performance of Sequential Apriori Technique}

In this subsection, we experimentally determine the accuracies of the maximal frequent patterns generated by the sequential apriori technique using the sessions reconstructed by different session reconstruction heuristics.  After the reconstruction of the sessions, they are processed by the sequential apriori algorithm in order to discover the frequent patterns in these sessions. Sequential apriori technique is also applied to the actual sequences generated by the agent simulator. Since, we know the frequent maximal patterns of the sequences of the agent simulator ($MP_{A}$), which correspond to the correct frequent patterns, we can determine the accuracies of different heuristics ($A_{H}$ stands for the accuracy of a heuristic H) by using the maximal frequent patterns generated by these heuristics ($MP_{H}$) as follows:

\begin{eqnarray}
A_H  = \frac{{|MP_A  \cap MP_H |}}{{|MP_A |}}
\end{eqnarray}

In our experiments, we have studied the accuracy by varying 4 different parameters, namely STP, LPP, NIP and the support. In each experiment we have fixed the three parameters (STP, LPP, and NIP) which are used to define the behavior of an agent, and obtained the accuracies for varying support values. For each one of these parameters, we have used two typical values, namely 0.10 and 0.20 for STP and 0.20 and 0.40 for LPP and NIP. Also, in each experiment the support is defined from 0.05\% to 0.25\%. Table 5 summarizes the parameters used in these experiments. Among these parameters the 4th experiment gave the lowest and the 5th experiment gave the highest accuracy results for all heuristics and support values. 

The important result of these experiments is the large improvement of the accuracies of the frequent pattern sessions in the second phase. As it can be seen from Figures 5 and 6, corresponding to the $4^{th}$ and the $5^{th}$ experiments, respectively, the accuracy of the second phase is always much higher than the accuracy of the reconstructed sessions. It is also observed that the accuracy of discovered frequent maximal patterns is about 30\% higher when Smart-SRA is used.  Similar results were also obtained for other 6 experiments.

\begin{table}
\centering
\caption{Parameters Used for Modeling Web Users}
\begin{tabular}{|c|c|c|c|} \hline
{\textbf{Experiment No}} & {\textbf{STP}} & {\textbf{LPP}} & {\textbf{NIP}} \\ \hline
{1} & {0.10} & {0.20} & {0.20} \\ \hline
{2} & {0.10} & {0.20} & {0.40} \\ \hline
{3} & {0.10} & {0.40} & {0.20} \\ \hline
{\textbf{4}} & {\textbf{0.10}} & {\textbf{0.40}} & {\textbf{0.40}} \\ \hhline{|-|-|-|-|}
{\textbf{5}} & {\textbf{0.20}} & {\textbf{0.20}} & {\textbf{0.20}} \\ \hline
{6} & {0.20} & {0.20} & {0.40} \\ \hline
{7} & {0.20} & {0.40} & {0.20} \\ \hline
{8} & {0.20} & {0.40} & {0.40} \\ \hline
\end{tabular}
\end{table}

\begin{figure}
	\centering
		\includegraphics{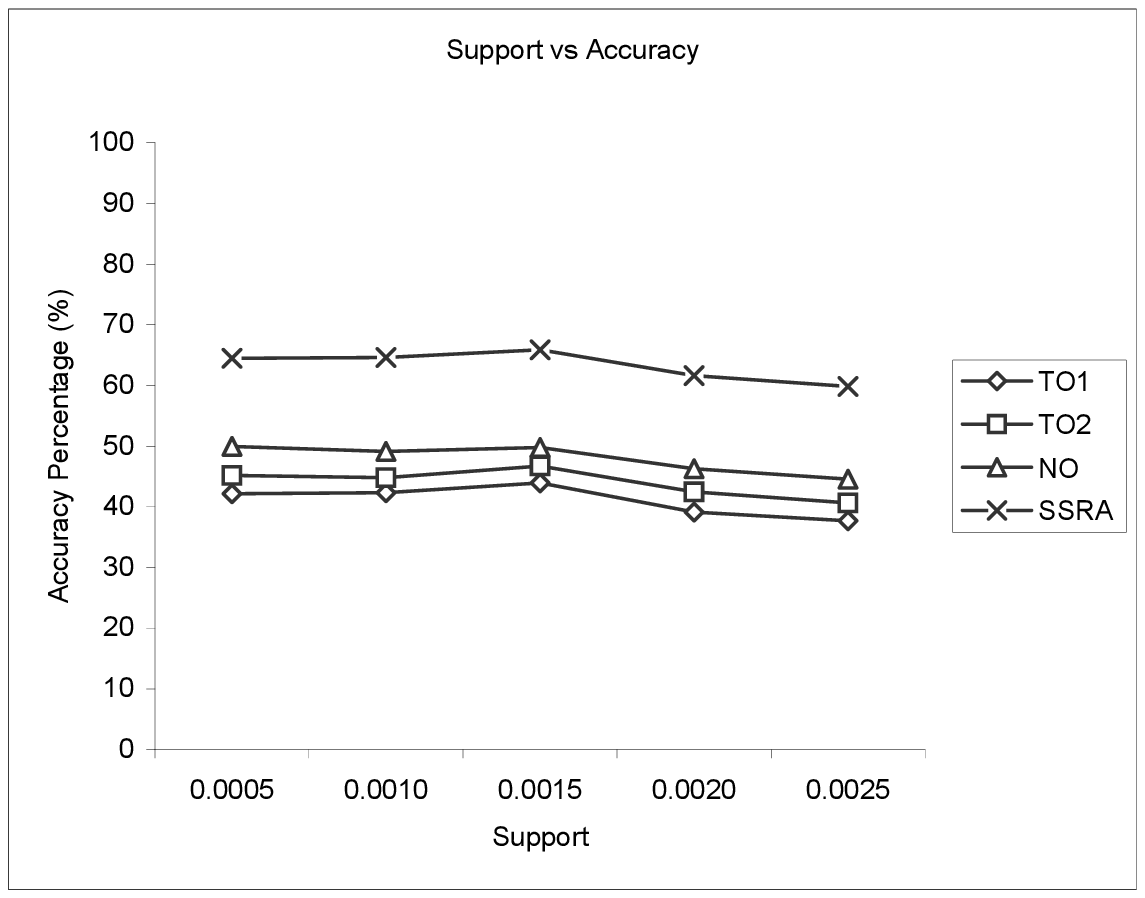}
	\caption{Support vs. Accuracy (Experiment no 4)}
\end{figure}

\begin{figure}
	\centering
		\includegraphics{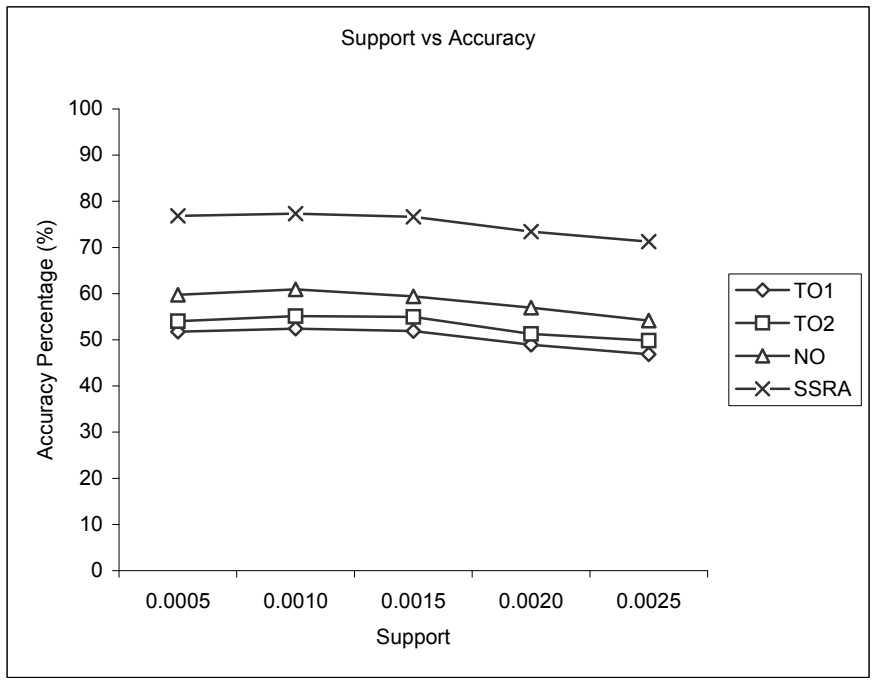}
	\caption{Support vs. Accuracy (Experiment no 5)}
\end{figure}

This result is not surprising because in the first phase real sessions must appear as sub-sessions of reconstructed sessions in order to be considered as correct, but, on the other hand, in the second phase, only frequent real patterns must appear as sub-sessions of frequent reconstructed sessions in order to be considered as correct. Thus, it is more likely to obtain higher accuracy in the second phase. In the session reconstruction phase, for 100\% accuracy of any pattern, it must appear in the reconstructed sessions for each of its occurrence in the actual sessions. On the other hand, for frequent pattern discovery phase, it is sufficient if the pattern appears as many times as it is required by the support value. Therefore, for frequent patterns we obtain much higher accuracies.  

\section{Conclusion and Future Work}

In this paper we have introduced a new frequent web usage pattern discovery method. Frequent patterns are discovered among the reconstructed sessions. Sessions can be reconstructed by using various heuristics. In our experiments we have used a recently developed heuristic Smart-SRA and time and navigation oriented heuristics for this first phase. Then, we have used a newly proposed sequential apriori technique in order to discover frequent patterns in the set of reconstructed sessions.

In the session reconstruction phase, for a reconstructed session to be assumed as accurate, it must include a session generated by the agent simulator. A frequent pattern is accepted as accurate if it appears only as a frequent pattern in the reconstructed sessions. Therefore, the accuracy increases after the pattern discovery phase, since complex navigational behaviors, which are hard to discover, but also infrequent are eliminated at this phase. 

The main purpose of WUM is to extract useful user navigation patterns. Therefore, it is not sufficient only to reconstruct user sessions from server logs. Capturing frequent user navigation patterns is more significant. This is achieved by employing frequent pattern discovery techniques after sessions are reconstructed. It is important to observe that the success of session reconstruction phase of WUM affects the success of the frequent pattern discovery phase. Moreover, our experiments also show that regardless of which heuristic is used for session reconstruction, in frequent pattern discovery phase the accuracy always increases. Also, by adjusting the support parameter of the apriori technique it is possible to control the frequency requirement of the common patterns searched in user navigation behaviors as well as the number of patterns discovered.

As a future work, modifying SRA heuristic for proactive session reconstruction can be considered by using other additional information sources. Also, our agent simulator can be improved in order to represent user navigation behaviors more correctly by adding new features.

\bibliographystyle{abbrv}
\bibliography{paper-1}

\begin{thebibliography}{10}

\bibitem{Nanopoulous01}
Y.~M. A.~Nanopoulos, D.~Katsaros.
\newblock Effective prediction of web-user accesses: A data mining approach.
\newblock In {\em WEBKDD}, 2001.

\bibitem{AgrawalS94}
R.~Agrawal and R.~Srikant.
\newblock Fast algorithms for mining association rules in large databases.
\newblock In {\em VLDB}, pages 487--499, 1994.

\bibitem{AgrawalS95}
R.~Agrawal and R.~Srikant.
\newblock Mining sequential patterns.
\newblock In {\em ICDE}, pages 3--14, 1995.

\bibitem{Bayir06-2}
M.~A. Bayir.
\newblock A new reactive method for processing web usage data.
\newblock Master's thesis, Middle East Technical University, 2006.

\bibitem{Bayir06-1}
M.~A. Bayir, I.~H. Toroslu, and A.~Cosar.
\newblock A new approach for reactive web usage data processing.
\newblock In {\em ICDE Workshops}, page~44, 2006.

\bibitem{CooleyMS97}
R.~Cooley, B.~Mobasher, and J.~Srivastava.
\newblock Web mining: Information and pattern discovery on the world wide web.
\newblock In {\em ICTAI}, pages 558--567, 1997.

\bibitem{CooleyMS99}
R.~Cooley, B.~Mobasher, and J.~Srivastava.
\newblock Data preparation for mining world wide web browsing patterns.
\newblock {\em Knowl. Inf. Syst.}, 1(1):5--32, 1999.

\bibitem{CooleyTS99}
R.~Cooley, P.-N. Tan, and J.~Srivastava.
\newblock Discovery of interesting usage patterns from web data.
\newblock In {\em WEBKDD}, pages 163--182, 1999.

\bibitem{Frias-MartinezK02}
E.~Frias-Martinez and V.~Karamcheti.
\newblock A customizable behavior model for temporal prediction of web user
  sequences.
\newblock In {\em WEBKDD}, pages 66--85, 2002.

\bibitem{FuS02}
Y.~Fu and M.-Y. Shih.
\newblock A framework for personal web usage mining.
\newblock In {\em International Conference on Internet Computing}, pages
  595--600, 2002.

\bibitem{Gaul00}
W.~Gaul and L.~Schmidt-Thieme.
\newblock Mining web navigation path fragments.
\newblock In {\em In Proceedings of the Workshop on Web Mining for E-Commerce},
  2000.

\bibitem{Gunduz03}
S.~G{\"u}nd{\"u}z and M.~T. {\"O}zsu.
\newblock A web page prediction model based on click-stream tree representation
  of user behavior.
\newblock In {\em KDD}, pages 535--540, 2003.

\bibitem{Pitkow99}
P.~P. J.~E.~Pitkow.
\newblock Mining longest repeating subsequences to predict world wide web
  surfing.
\newblock In {\em USENIX}, 1999.

\bibitem{MobasherCS00}
B.~Mobasher, R.~Cooley, and J.~Srivastava.
\newblock Automatic personalization based on web usage mining.
\newblock {\em Commun. ACM}, 43(8):142--151, 2000.

\bibitem{MobasherDLN02}
B.~Mobasher, H.~Dai, T.~Luo, and M.~Nakagawa.
\newblock Discovery and evaluation of aggregate usage profiles for web
  personalization.
\newblock {\em Data Min. Knowl. Discov.}, 6(1):61--82, 2002.

\bibitem{NasraouiK02}
O.~Nasraoui and R.~Krishnapuram.
\newblock An evolutionary approach to mining robust multi-resolution web
  profiles and context sensitive url associations.
\newblock {\em International Journal of Computational Intelligence and
  Applications}, 2(3):339--348, 2002.

\bibitem{PeiHMZ00}
J.~Pei, J.~Han, B.~Mortazavi-Asl, and H.~Zhu.
\newblock Mining access patterns efficiently from web logs.
\newblock pages 396--407, 2000.

\bibitem{PierrakosPPS03}
D.~Pierrakos, G.~Paliouras, C.~Papatheodorou, and C.~D. Spyropoulos.
\newblock Web usage mining as a tool for personalization: A survey.
\newblock {\em User Model. User-Adapt. Interact.}, 13(4):311--372, 2003.

\bibitem{SchechterKS98}
S.~E. Schechter, M.~Krishnan, and M.~D. Smith.
\newblock Using path profiles to predict http requests.
\newblock {\em Computer Networks}, 30(1-7):457--467, 1998.

\bibitem{ShahabiK03}
C.~Shahabi and F.~B. Kashani.
\newblock Efficient and anonymous web-usage mining for web personalization.
\newblock {\em INFORMS Journal on Computing}, 15(2):123--147, 2003.

\bibitem{Spiliopoulou00}
M.~Spiliopoulou.
\newblock Web usage mining for web site evaluation.
\newblock {\em Commun. ACM}, 43(8):127--134, 2000.

\bibitem{SpiliopoulouF98}
M.~Spiliopoulou and L.~Faulstich.
\newblock Wum - a tool for www ulitization analysis.
\newblock In {\em WebDB}, pages 184--103, 1998.

\bibitem{SpiliopoulouMBN03}
M.~Spiliopoulou, B.~Mobasher, B.~Berendt, and M.~Nakagawa.
\newblock A framework for the evaluation of session reconstruction heuristics
  in web-usage analysis.
\newblock {\em INFORMS Journal on Computing}, 15(2):171--190, 2003.

\bibitem{SrikantY01}
R.~Srikant and Y.~Yang.
\newblock Mining web logs to improve website organization.
\newblock In {\em WWW}, pages 430--437, 2001.

\bibitem{SrivastavaCDT00}
J.~Srivastava, R.~Cooley, M.~Deshpande, and P.-N. Tan.
\newblock Web usage mining: Discovery and applications of usage patterns from
  web data.
\newblock {\em SIGKDD Explorations}, 1(2):12--23, 2000.

\end{thebibliography}

\end{document}